# DEEPFAKES AND HIGHER EDUCATION: A RESEARCH AGENDA AND SCOPING REVIEW OF SYNTHETIC MEDIA

Jasper Roe [1*], Mike Perkins [2]

[1] James Cook University Singapore, Singapore
[2] British University Vietnam, Vietnam.

* Corresponding Author: jasper.roe@jcu.edu.au

## Abstract

The availability of software which can produce convincing yet synthetic media poses both threats and benefits to tertiary education globally. While other forms of synthetic media exist, this study focuses on deepfakes, which are advanced Generative AI (GenAI) fakes of real people. This conceptual paper assesses the current literature on deepfakes across multiple disciplines by conducting an initial scoping review of 182 peer-reviewed publications.

The review reveals three major trends: detection methods, malicious applications, and potential benefits, although no specific studies on deepfakes in the tertiary educational context were found. Following a discussion of these trends, this study applies the findings to postulate the major risks and potential mitigation strategies of deepfake technologies in higher education, as well as potential beneficial uses to aid the teaching and learning of both deepfakes and synthetic media. This culminates in the proposal of a research agenda to build a comprehensive, cross-cultural approach to investigate deepfakes in higher education.

# Introduction

The pace of the development of Artificial Intelligence (AI) technologies has led to significant concern in many areas of society, including educational contexts. As a result, research agendas on Generative AI (GenAI) in tertiary education have been established (Lodge et al., 2023); however, to date, no review or research agenda has specifically focused on deepfakes in tertiary education. Deepfakes are GenAI outputs which comprise realistic audio, visual, or media outputs that depict false or inaccurate information (Akhtar, 2023). The major consequence of deepfakes is that they can portray an individual doing something or saying something that they have never done, marking an unprecedented shift in the ability to distort reality (Appel & Prietzel, 2022). As tertiary education institutions are centres of learning, the potential implications of such false information are highly important for students, teachers, and university leadership, thus warranting stakeholder attention.

While we primarily focus on deepfakes in this paper, it is also important to mention that non-deepfake synthetic media are also highly important in tertiary education, and this forms part of our proposed research agenda. For example, Microsoft's recently released VASA-1 framework can produce sophisticated human-like avatars with nuanced facial expressions and lip synchronisation through a single image. We distinguish between this form of synthetic media and deepfakes, as VASA-1 has been demonstrated only using synthetically generated visual identities with tools such as StyleGAN2 and DALL-E 3, rather than real people (Xu et al., 2024).

Our study aims to contribute to the literature on educational technology in tertiary education by addressing three distinct aims. The first is to conduct a rapid scoping review to assess the literature on deepfakes and synthetic media across multiple fields and to relate the impact of these findings to the context of tertiary education. The second aim is to identify potential strategies and best practices that can serve as a starting point for crafting coherent institutional approaches to dealing with deepfakes. The third aim is to propose a research agenda which establishes priority areas for exploring and addressing both the potential challenges and benefits of deepfake technology in higher education and synthetic media. Consequently, this article offers a novel contribution which serves to set the foundation for the study of deepfakes and synthetic media in education.

The term deepfake was coined in 2017 by a user on the social media site Reddit, to refer to a combination of the terms 'deep learning' and 'fakes' (Kietzmann et al., 2020), in contrast to 'shallow fakes' or 'cheap fakes' which revolve around simpler, less convincing methods of media manipulation, such as slowing videos or using photo editing software. While we use the English terminology 'deepfake' in this article, it is important to note that deepfakes are a global phenomenon, do not only occur in Western contexts, and have their own terms in other languages (e.g. huanlian in Chinese (de Seta, 2021)). Cases of deepfakes spreading on social media have been documented in Kenya (Mayoyo, 2023), India (Nema, 2021), Russia (Samoilenko & Suvorova, 2023), and China (de Seta, 2021).

Although deepfakes have been recognised since 2017, the underlying technology required to successfully produce these manipulated media outputs has only recently been sufficiently developed to become widely accessible (Busacca & Monaca, 2023). Public awareness of the issue has also grown through several recent high-profile scandals, including pornographic deepfakes of celebrities such as Taylor Swift (Saner, 2024). These cases have led to Reddit banning the original 'deepfake' subreddit in which the term was originally coined (Appel & Prietzel, 2022), and in the United Kingdom, creation of nonconsensual, explicit deepfake images was criminalised in April 2024, attracting fines and even jailtime for creators (Milligan, 2024).

Historically speaking, media and image manipulation is not new (Ali et al., 2021), and propagandist video manipulation has been well documented in WWII (Fallis, 2021). To this end, manipulation or doctoring of content in general has been called 'as old as the media industry itself' (Kietzmann et al., 2020). That said, deepfakes pose an entirely new challenge as they are highly accessible, easy, and quick to create, require little expertise, and can spread readily in online domains (Agarwal et al., 2019). It has been argued that society is not currently able to deal with the emergence of deepfakes (Karnouskos, 2020), and there is currently no common strategy in legal, political, and institutional domains to address them. Media outlets and internet service providers have idiosyncratic approaches to dealing with deepfakes (Vizoso et al., 2021). There are also growing concerns about deepfakes being used to create child sexual abuse material (CSAM) (Hern, 2024). Consequently, there is a growing consensus among political and legal stakeholders that deepfakes require some form of regulation (Hern, 2024, p. 24; Langa, 2021), lest the veracity of all audio and visual media come under question, a scenario labelled an 'infocalypse' (Schick, 2020). At the time of writing, no coherent legal and institutional framework is forthcoming; therefore, we contend that it is up to practitioners and researchers in educational technology to take the lead in addressing deepfakes in a tertiary education context, and that this represents a professional obligation to meaningfully engage with GenAI, so as not to let the agenda be controlled by other actors (Thompson et al., 2023).

## Reviewing The Landscape of Deepfakes and Deepfake Research

Deepfakes are produced through deep learning procedures which typically involve Generative Adversarial Networks (GANs) or more recently, advanced diffusion models (Appel & Prietzel, 2022). Historically, deep learning to produce visual materials has required large sets of training data. This is one reason why celebrities have been common targets, as huge libraries of images and video material on these individuals already exist (Kietzmann et al., 2020). More recently, end-to-end systems which generate videos of talking heads using only a single still image and an audio clip have been developed (Vougioukas et al., 2019; Xu et al., 2024), suggesting that the ability and development of synthetic media and deepfakes will continue to improve and that deepfakes of anyone with a single image may be achievable. Similar to large language models (LLMs) and GenAI image generators, deepfake technologies are increasingly accessible to the general population. There are multiple open source applications for deepfake creation, including DeepFaceLab, FaceApp, Wombo, Zao, and FaceApp (de Rancourt-Raymond & Smaili, 2022). One of the most popular applications, FakeApp, makes the creation of deepfakes possible for users with basic technological capabilities quickly and easily (Ali et al., 2021). Therefore, deepfakes are expected to become more common in the near future.

There are no unified or agreed upon common categories for deepfakes, but there are areas of commonality across definitions and frameworks. Farid (2022) outlines three categories of visual deepfake. The first is known as 'face-swap'. A common example of this deepfake is swapping the face of an actor into another movie in which they did not appear. The second category is lip synchronisation (lip-sync), in which the modification of a video aims to match the speaker's lips with audio recording. The third category is 'puppet-master' and requires the animation of a person's image by another performer (the 'puppet'). Akhtar (2023) suggested four alternate types of visual deepfake manipulation: swapping of identities, face reenactment, manipulation of attributes, and synthesis of the face in its entirety. Kietzmann et al. (2020)'s taxonomy of deepfakes extends beyond visual media, encompassing photo deepfakes which involve face and body swapping, audio deepfakes which include using voice-swapping, and video and combined video and audio deepfakes. Regardless of these minor differences in classification, deepfakes of all varieties produce similar output: believable, realistic media which is predicated on some aspect of 'fakery' or inaccurate representation, thus distinguishing it from synthetic media which can use a non-existent or artificial identity.

While our literature review aims to understand the scope of deepfake and synthetic media research from the perspective of tertiary education, there have been many broader reviews of the deepfake literature in other disciplines. Bode et al. (2021) points out that there has been an 'explosion' of multidisciplinary research on deepfakes, often focusing on erosion of truth and trust, the creation of fake pornography, and political manipulation. A systematic review of deepfakes identified that much of this research is within the fields of computer science, politics, and law (Godulla et al., 2021), while a bibliometric analysis of 331 research articles surrounding deepfakes on Web of Science and Scopus databases found similar patterns (Gil et al., 2023). In conducting our own assessment of the literature, we undertook a search on the Scopus database for the terms 'deepfake', 'GAN, ', synthetic media', 'fake media', and 'AI generated media' with the co-operators 'education', 'higher education' and 'tertiary education'.

Given that we aimed to ascertain initially whether there was research on deepfakes in tertiary education, we adopted a scoping review procedure (Mak & Thomas, 2022). We obtained 182 peer-reviewed journal publications, conference proceedings, and chapters in edited books regarding deepfakes and synthetic media, but no works that could be described as specifically focusing on the tertiary education context. This indicated that there was not sufficient literature for a detailed review on deepfakes in education, although there is limited existing literature regarding deepfake education among younger cohorts (Ali et al., 2021; Blankenship, 2021; Murillo-Ligorred et al., 2023). After assessing and reviewing the literature we identified three broad trends that characterize current scholarship on deepfakes and synthetic media, and confirm the findings of Godulla et al. (2021) and Gil et al. (2023). The trends we identified comprise deepfake detection, malicious applications of deepfake technologies, and positive applications of deepfake technologies. In the below, we provide an overview of these findings, beginning with the detection of deepfake technology, followed by applications. We divide current applications into two sections, based on an evaluative judgement of the potential impacts as being either negative (i.e. causing harm or societal and individual damage) or positive (i.e. bringing benefits to society or the individual).

**Detection of Deepfakes**

Our review found that the majority of literature on deepfakes focuses on the detection and evasion of deepfake content, mirroring current scholarship on detection of GenAI text (Anderson et al., 2023; Perkins et al., 2023; Weber-Wulff et al., 2023). Research has shown that while less sophisticated, early deepfakes were more detectable to the naked eye as a result of inconsistencies, such as blinking frequency (Agarwal et al., 2019), more recent iterations of deepfake images are difficult for humans to detect (Chadha et al., 2021). One bilingual study on audio deepfake detection in both Mandarin and English (N=529) found that listeners only correctly identified deepfakes 73% of the time, regardless of language (Mai et al., 2023). This is concerning as the results of other investigations suggest that people may overestimate their ability to detect deepfakes (Köbis et al., 2021).

Given the difficulties in relying on human-led detection of deepfakes, it is unsurprising that computational methods of detection have become a popular topic of enquiry. To this end, it seems that there is a 'cat and mouse game' of detection and creation (de Seta, 2021), or an arms-race scenario, which has been noted in other forms of high-technology detection processes in an AI-driven higher education landscape (Roe & Perkins, 2022). Many studies continue to posit novel detection techniques to improve accuracy of deepfake detection (Giudice et al., 2021; He et al., 2021; Hernandez-Ortega et al., 2020; Huang et al., 2020; Jung et al., 2020; Porcile et al., 2024; Qi et al., 2020; Tolosana et al., 2020). However, some argue that as a result of this tension between identification and evasion, deepfakes are not reducible to a problem which is solvable by detection (Jacobsen & Simpson, 2023).

**Adverse Impacts Of Deepfake Technologies**

Following deepfake detection, the negative potential impacts of deepfake technologies seem to account for much of the extant literature outside of the domain of computer science. These include the dangers of deepfake misinformation on political processes and elections (Diakopoulos & Johnson, 2021; Dobber et al., 2021; Hameleers et al., 2022; Vaccari & Chadwick, 2020), pornography (Arhiptsev et al., 2021; Burkell & Gosse, 2019; Delfino, 2019; Karasavva & Noorbhai, 2021; Öhman, 2020) legal and ethical aspects of deepfakes (Citron & Chesney, 2019; Franks & Waldman, 2018; Meskys et al., 2019) and the philosophy of deepfakes in moral terms (de Ruiter, 2021). A common theme that recurs in describing the risks of deepfakes is the ability for actors to take advantage of the 'liar's dividend' i.e. the ability for individuals to deny factual content as fake (Ahmed, 2023).

One of the dominant concerns of deepfake disinformation in social and political contexts is the ability for it to sow social divisions (Harris, 2021), which can take place along the lines of political affiliation, gender, race, and class, while simultaneously lowering trust in institutions and authorities (Helmus, 2022). As tertiary educational institutions are tasked with providing education and care for students, this is also of great potential impact. Social discontent, instability, and lowering of trust in the institution are all specific effects of deepfakes which may be amplified within the context of a university or college campus.

Negative impacts also extend to the individual. Voice authentication of secure information such as bank accounts is no longer viable with the advent of vocal deepfakes (OpenAI, 2024), and deepfake technology can be used to produce non-consensual media about a person or persons, which can invade privacy, damage individuals and families or be used for identity theft (de Rancourt-Raymond & Smaili, 2022). In relation to non-consensual pornography, this is now an ongoing issue that may affect students or staff in a higher education context. In 2019, a $50 app named DeepNude allowed users to digitally 'undress' anyone from an image, and now multiple applications exist for this purpose (Farid, 2022). This can take the form of cyber-bullying, which may be specifically used to objectify and degrade women (Burkell & Gosse, 2019). In one study on social media use and deepfakes in Kenya, a majority of respondents (N=978) had been victims of sexual violence on social media, which included being victims of deepfakes (Mayoyo, 2023). Moreover, as legal regulations and procedures surrounding deepfakes in much of the world have not yet developed, there is little opportunity to achieve justice (Kietzmann et al., 2020).

**Beneficial Impacts Of Deepfake Technology**

Our review also found that there are beneficial applications of deepfake technology, with some of these being relevant to tertiary education. Deepfakes can fulfil an entertainment function, and be a source of joy, fun, and innocent trickery (Pashentsev, 2023), confirmed by the fact that entire channels exist on YouTube for deepfake parody videos, garnering millions of views (Langguth et al., 2021). Deepfakes and synthetic media have the potential to efficiently bolster media production, having already appeared in feature films, and in educational campaigns against public shootings (Farid, 2022). Deepfakes and synthetic media may have uses in advertising (Campbell et al., 2022) e-commerce and branding (Bode et al., 2021), which may be of use to educational institutions. Deepfakes have also been used as a tool of public communication through artwork. Cheres & Groza (2023) produced an AI-based at installation to raise awareness of the threats of social media, by using deepfake technology to highlight the way in which users are providing substantial training data to AI-models through uploading images of their likeness.

In our scoping review, we found only one article which specifically describes an existing (rather than theoretical) positive use case for deepfakes in an educational context, albeit in the field of tourism. In this work, Kwok & Koh (2021) point out that the Salvador Dali Museum in Florida, USA, uses a deepfake of the artist, allowing for a personalised approach in which Dali educates visitors about his artwork himself. As sites of research and scientific enquiry, tertiary educational institutions can play a key role in promoting and advocating for these positive uses of deepfakes. This could include bringing deepfake technology to historical characters, which could be combined with Virtual Reality (VR), which has been shown to give students realistic educational experiences (Calvert & Hume, 2022). In the non-academic literature, OpenAI notes potential benefits of synthetic media in the form of artificially generated voices. This could include helping patients recover their voices (which could involve a deepfake), improving the delivery of health information in remote settings, and provide reading assistance to non-readers (OpenAI, 2024).

The findings of this scoping review reveal that there are both beneficial and adverse impacts of deepfakes and synthetic media which are relevant to higher education, university students, staff, and stakeholders. However, the lack of research in these contexts signifies a gap, which our research agenda intends to address. At the same time, while our review is comprehensive and confirms similar results to that of Godulla et al. (2021) and Gil et al. (2023), it is limited by the scoping review's lack of formal quality assessment of included studies. Nevertheless, the findings from this review give an account of the current literature on deepfakes and synthetic media as it pertains to education. Following this section, we address our second aim by specifically identifying how some of these adverse impacts may apply to university stakeholders.

## Risks and Potential Applications of Deepfake Technology in HE

### Cyberbullying Of Students Or Faculty

Cyberbullying is prevalent in universities, with one study finding that 38% of North American students (N=439) knew a victim of cyberbullying (MacDonald & Roberts-Pittman, 2010). Although the definition of cyberbullying may vary cross-culturally (Akbulut & Eristi, 2011), core elements include online harassment, threats, circulating embarrassing or sensitive information, and dissemination of visual media, such as videos of bullying instances, all of which can have serious consequences on mental health (Kowalski et al., 2014).

Our literature review revealed that fraud and non-consensual pornography are two negative impacts of deepfake technology; as such, core cyberbullying elements may involve the use of deepfakes (Langguth et al., 2021). Cyberbullying deepfakes may include creating videos of an individual making a claim which is controversial or damaging via audio or video, or through sexual bullying by creating non-consensual pornography, as has been noted in the Kenyan university context (Mayoyo, 2023). In university settings, cyberbullying may have immediate career implications for students upon graduation (Cunningham et al., 2015), making it an especially prominent risk. As accessibility and barriers to the entry of deepfake technology continue to decrease, and the convincing nature of the technology increases, cyberbullying deepfakes become an ever-more likely scenario, meaning that institutions must proactively find ways to address it.

### Academic Dishonesty

Initial concerns regarding AI and academic dishonesty revolved around the detection of AI-generated text and the use of GenAI tools to violate the values of academic integrity (Cotton et al., 2023; Perkins, 2023). However, deepfakes pose equal risk to academic integrity. Although low-technology methods of manipulating research and result data have occurred for centuries, the use of deepfake and AI techniques to conduct image-based fraud may make it more difficult to detect (Sundar et al., 2021). For example, deepfake technologies may be used to create fake data, manipulate research findings, or distribute false material to impact the success of other students. The risk of this is great, given the increasing pressure on students and academics in a competitive and instrumental system of education (Roe, 2022).

Among a variety of education stakeholders (including undergraduate and graduate students and educators), Doss et al. (2023) found that between 27% and 50% of participants could not identify deepfake videos, meaning that believable, fraudulent videos, or other deepfake materials could impact academic honesty and integrity. Deepfakes can even extend to admission fraud, being used to create fake application material, audio or video testimonials, or recommendations, thus impacting equity and inclusion in higher education. Diploma mills which mimic established universities continue to threaten the validity and integrity of qualifications (Roe & Perkins, 2023), and deepfakes could conceivably be used to create false video testimonials or endorsements from high-profile academics or institutions.

**Decaying Institutional Trust And Reputation**

Deepfakes can propagate disinformation and thus lead to the acquisition of false beliefs (Fallis, 2021). In a higher education context, deepfakes could be produced regarding university leadership, staff, or students, in which they endorse or espouse controversial or untrue beliefs. This could lead to social divisions on campus and lower trust in the university as an institution and authority. Furthermore, younger students are vulnerable to misinformation (Doss et al., 2023), young adults and adolescents may be more impressionable as a result of social factors (Gwon & Jeong, 2018), and so are prime targets for disinformation campaigns (Ali et al., 2021). Consequently, the risk for higher education students to be affected by deepfakes in a way that damages the reputation of the institution or community spirit may be larger than in the general population.

In a highly marketised competitive landscape, reputation plays a vital role in universities (Angliss 2022). As the main risks of deepfakes to businesses are reputation and trustworthiness (Mustak et al., 2023), such risks may also affect the operations, aspirations, and financial success of higher education institutions (HEIs). Such reputational damage may be irreparable (Moerschell & Novak, 2020) and should be treated seriously by university leadership.

**Strategies For Countering The Risks Of Deepfakes In Higher Education**

Overall, in addressing our second aim, we contend that the greatest potential threats to tertiary education stakeholders regarding deepfakes are cyberbullying, reputational damage, and academic dishonesty. As educational institutions do not yet have effective ways of preparing students to tackle the issues of deepfakes and disinformation (Naffi et al., 2023), we propose several broad strategies based on the extant literature in the fields of cyberbullying and crisis management for institutions to begin a proactive approach to countering the potential threat of deepfakes.

**Augment Cyberbullying Policies**

Research has shown that many university policies on codes of conduct and discipline do not explicitly refer to cyber behaviours (Faucher et al., 2015). With this in mind, a code of conduct or anti-cyberbullying policy should refer to specific behaviours (for example, creating non-consensual explicit deepfakes or spreading disinformation). It is necessary for such policies to be specific and not overbroad to avoid infringing on students' rights (O'Connor et al., 2018). There is no universal approach that can be taken with deepfake or cyberbullying policies, given that tertiary institutions operate in cultural and geographic contexts which may have differing approaches to conflict resolution and views on the legality or permissibility of deepfakes. In the U.K., for example, recent changes in legislation have criminalised the production of non-consensual sexually explicit deepfakes (Milligan, 2024), so institutional policies may be developed in line with legal reporting standards. This may not be possible in all areas globally, but regardless of the context, taking a preventative and proactive approach based on the specific context of the institution is necessary (O'Connor et al., 2018), and anonymous online reporting processes may reduce the risk of retaliation (Cunningham et al., 2015). Given the gendered nature of deepfakes and the fact that cyberbullying often affects women more frequently at the university level (Faucher et al., 2015), gender-specific policies should be explored.

**Undertake Educational Interventions**

Educational interventions may be undertaken to inform about the risks of deepfakes and synthetic media. Deepfakes may be specifically included as a component in anti-cyberbullying awareness interventions, given that education and training can be effective in reducing the prevalence of cyberbullying (Li, 2007). However, educating stakeholders about deepfakes in general may also be beneficial, as it can be part of a broader media literacy effort in an age of growing disinformation. In one of the few studies in this area, Ali et al. (2021) described a series of activities to educate middle-school students about how deepfakes are created and identified as a method of fostering critical literacy. Such educational programs could also be valuable for university students and academic staff.

By undertaking deepfake education, students and faculty can be taught what deepfakes are and how to suspect them. This can help to 'pre-bunk' or 'inoculate' students and staff to be sensitive and aware of deepfakes as a misinformation strategy (Horvitz, 2022). Sharing some of the more relevant research findings on the psychology of deepfake impacts, such as the fact that deepfakes may even alter memories of events (Murphy et al., 2023), may also be beneficial in highlighting the potential consequences of deepfake disinformation.

In summary, by exposing students and staff to facts through targeted educational interventions and training programs, higher education stakeholders can develop better resilience against deepfake technologies and learn to 'put their guard up' against such new threats (Hancock & Bailenson, 2021). In an effort to combat educational fraud, such interventions may also promote trust, integrity, and transparency in the higher education process. As educational interventions like this have previously been shown to be effective in reducing instances of academic misconduct (Perkins et al., 2020), ensuring students are familiar with the potential damaging influences of deepfakes may reduce misuse of the technology.

**Develop Crisis Management Plans**

Higher education institutions often treat crises as rare occurrences, making them ill-equipped to respond (Booker Jr., 2014). In reference to a deepfake 'crisis' in which there is an immediate and high level of risk to the integrity of the institution, developing a specific management and communication plan is necessary. For example, a deepfake crisis may include the production of fabricated content of an academic leader saying something inflammatory or insensitive, or show them doing something that would cause reputational damage. Such disinformation may spread quickly via social media, and a coordinated plan to counter it may focus on media strategies aimed at limiting and repairing reputational damages. Such plans can also be proactive, including the development of communication strategies to inform people who are at risk of being targeted as well as other stakeholders and the media (Moerschell & Novak, 2020).

Building on Moerschell and Novak (2020), several examples of crisis management planning can be adapted to deal with deepfake incidents. First, practical considerations can involve developing a 'dark site' to be published only in the event of a crisis to communicate with stakeholders, which may enable institutions to combat any disinformation spread by a deepfake and provide accurate information to the recipients. A social media page can also be developed to manage crises, including deepfake crises, act against conflicting social media information, and release information quickly and strategically (Moerschell & Novak, 2020). Fundamentally, crisis management plans for deepfakes depend on the information contained in the falsified content. However, proactive communication planning and crisis management may mitigate reputational damage to institutions, students, and staff involved. Given that the effectiveness of organisational responses to disinformation and fake news varies depending on several factors (Vafeiadis et al., 2019), formulating a set of protocols for deepfake crisis management is an important research task.

**Potential Applications of Synthetic Media Technology in HE**

Although the risks of deepfake technology are significant and warrant serious consideration, there are potential positive applications of the underlying technology in higher education that should not be overlooked. In these more positive aspects, we refer to the broader category of synthetic media production. One promising use of this technology is the creation of low-cost, high-quality educational videos. By employing readily available tools which can create video scripts from research papers or lecture slides, complete with naturalistic errors and variations, and integrating it with avatar-based videos, universities can produce additional educational content more efficiently and cost-effectively than relying on faculty alone. This approach could ease the burden on instructors, freeing up their time for other essential teaching and research activities, while still providing students with engaging and consistent learning materials.

Deepfake-powered video creation can also make educational content more inclusive and accessible. Academic staff who are less comfortable being on camera, or who face language barriers, could create compelling videos by scripting the content and having it delivered in their own voice and image through deepfake technology. This may help a wider range of educators to create video content that connects with their students. Similarly, universities could harness deepfakes to create video messaging featuring senior leadership or guest experts who have limited time to record content in person. For instance, a university president could virtually 'deliver' a commencement speech crafted by speechwriters, or an influential thought leader could 'present' at a conference without needing to be physically present. These 'synthetic guest speakers' may increase access to valuable insights and perspectives that would not otherwise have been possible to obtain. The emotional impact of feeling addressed by a renowned expert 'in person' could enhance engagement with the content.

Another potential application is the creation of immersive educational simulations. For example, historical events could be brought to life by generating synthetic videos of key figures, allowing students to experience a sense of direct interaction. Alternatively, complex scientific concepts could be explained by virtual instructors, who are able to display physically impossible feats that demonstrate the ideas being taught.

However, the use of deepfakes in educational contexts involves significant ethical considerations and risks. It is crucial that universities be fully transparent about where and how they employ synthetic media. Students must be educated to critically evaluate the credibility and quality of AI-generated content, not simply accept it at face value. As Bearman et al. (2024) argue, building students' evaluative judgment and media literacy capabilities is essential in a world where misleading deepfakes are increasingly prevalent. Strict ethical guidelines and oversight will be needed to ensure that deepfake technology in education is being used responsibly in ways that augment rather than undermine learning, and further research is needed to test the acceptability and impact of these use cases.

## Deepfake Education Research Agenda

Academic research plays a vital role in advancing the development of deepfake technology (Farid, 2022), as evidenced by the dramatic expansion of journal articles on this topic over the past few years (Gil et al., 2023). However, much of this research focuses heavily on the detection, creation, and political dimensions of deepfakes. There is less emphasis on the educational and pedagogical value of deepfakes or on how to safeguard students, institutions, and faculty from their effects. Consequently, we call for greater enquiry into several aspects of deepfake use within educational populations and contexts. This research should be interdisciplinary in nature (Akhtar, 2023), bringing together insights from computer science, social science, education, ethics, and other relevant fields to fully understand the complex implications of deepfakes in higher education.

**Faculty And Student Perceptions And Vulnerabilities Across Contexts**

Our review found that most cultural and political research on deepfakes is U.S.-centric (Ali et al., 2021); however, the risks and implications for education are global. There is a need for more studies exploring how academic staff and students in different cultural settings perceive AI-generated media as well as their ability to distinguish it from authentic content. Differences in digital literacy, socioeconomic factors, and information ecosystems may shape student vulnerability to deepfakes. Establishing a baseline of current levels of deepfake knowledge among learners worldwide would provide a helpful starting point for designing targeted interventions to increase deepfake literacy.

Research should examine how deepfakes may disproportionately impact marginalised groups. Deepfake technology raises the risk of targeted attacks and disinformation campaigns against women, minorities, and other underrepresented populations in academia. Understanding these unique vulnerabilities is crucial for developing support services and institutional policies that promote a safe and inclusive learning environment.

**Educational Initiatives To Build Deepfake Literacy And Resilience**

With concerns growing about student and educator readiness to operate in an educational environment rife with synthetic media, there is a need to evaluate the effectiveness of different approaches to deepfake literacy education. While some efforts are being made to educate younger students about deepfakes (Ali et al., 2021; Blankenship, 2021), empirical studies should test and compare specific pedagogical strategies for building university students' and staff critical evaluation skills. This is particularly important given the broader challenges faced by academia regarding the detection of text-based GenAI content (Chaka, 2023, 2024; Perkins et al., 2023, 2024; Weber-Wulff et al., 2023). Research could also explore how deepfake detection training, metaliteracy education, and other related interventions impact learners' ability to identify and resist manipulated content. Establishing best practices for deepfake-focused digital literacy will be essential for both classroom learning and campus awareness campaigns, especially as GenAI-supported video production becomes more commonplace and integrated into commercial products, such as Adobe's suite of video editing software (Weatherbed, 2024).

Beyond explorations of literacy about this technology, research should consider how to cultivate psychological resilience to the distressing effects of deepfakes. Exposure to disturbing deepfakes might inflict serious mental health harm. Therefore, investigating strategies to protect students against these impacts and providing support to affected individuals should be a priority. Research could also explore the potential for educational deepfakes to be used for learning purposes as previously discussed, or to build resilience to political messaging.

**Proactive And Responsive Institutional Practices**

As deepfakes become more prevalent in society, posing risks of reputational damage, fraud, and social division, HEIs need robust plans and protocols to prevent and manage related incidents on campuses. Case studies of institutions that have dealt with deepfake-related issues could provide valuable lessons and best practices for effective response. Research of this nature could inform the development of broader guidelines and principles that shape decisions regarding stakeholder communications, support for affected parties, investigative procedures, and coordination with external entities.

Institutions would also benefit from research on proactively reducing deepfake risks and harms, as well as exploring potential institutional use cases. This could include evaluating the effectiveness of different policy approaches, such as codes of conduct prohibiting deepfake misuse, mandatory training for students/staff, or disclosure requirements for educational deepfakes. Studies may also explore the role of campus initiatives, reporting systems, counter-messaging campaigns, and other prevention strategies. Establishing an evidence base for impactful practices will be key as institutions begin to create more comprehensive deepfake governance frameworks. Although evidence suggests that HEIs have been slow to adapt to the rapid growth of GenAI tools (Perkins & Roe, 2023), given the increasing institutional knowledge bases in these areas, institutions may now be more capable of producing effective deepfake policy, but require evidence to base this on. Ongoing work, such as the R.E.A.L framework for managing deepfake risks(Kietzmann et al., 2020), offers potential starting points for this.

**Ethical And Beneficial Applications For Teaching And Learning**

While much of the concern around deepfakes centres on their potential for harm, the technology may also hold promise for enhancing education when used responsibly. Therefore, researchers should examine how synthetic media can be leveraged to create engaging learning experiences, increase access to knowledge, and promote inclusivity. For example, studies might explore the pedagogical value of interactive deepfake simulations for bringing historical events to life or institutional benefits such as effective, low-cost messaging from leadership.

Realising these benefits will require a strong foundation of research on the ethical dimensions of educational deepfakes. The use of synthetic content in teaching and assessment raises complex questions about authenticity, consent, privacy, and intellectual property that need to be rigorously explored. Research should consider what disclosure and oversight mechanisms would be appropriate in different circumstances, how to handle student data used to generate content, and how to ensure that deepfakes do not mislead learners or undermine educational goals. Frameworks for implementing AI-generated content in ways that align with learning science principles and institutional values will therefore be needed to ensure that these use cases are carried out in an ethical manner.

**Long-Term And Interdisciplinary Approaches**

More research is needed on the long-range effects that deepfakes may have on the core aspects of HE. For instance, longitudinal studies could track how increased deepfake usage influences learning outcomes, student engagement, research quality, academic integrity, and campus discourse over time. Examining the interplay of deepfakes with other key issues shaping the future of higher education will also be important. Exploring these bigger-picture questions will help institutions understand their deepfake risks and responsibilities in context, and contribute to effective policy development.

As Akhtar (2023) argues, interdisciplinary research on deepfakes should be prioritised. Studies that bridge computer science and AI with social and political sciences are needed to fully understand the technical, psychological, and cultural dimensions of deepfakes in HE. The challenges posed by deepfakes cannot be addressed by any discipline alone; therefore, there is potential for the development of new crossovers between disciplines.

**Impacts On Individuals Affected By Deepfakes**

Finally, we call for research that takes a proactive approach to understand the specific impacts of those who are cyberbullied, victimised, or reputationally damaged by deepfake incidents. We echo the point made by Karasavva and Noorbhai (2021) that research on how to best support victims of deepfake pornography and other specific issues is warranted given that deepfakes are likely to become commonplace (Hancock & Bailenson, 2021) and continue to grow over time (Kietzmann et al., 2020).

# Conclusion

Deepfakes are not the first form of digital manipulation and will not be the last (Farid, 2022). To tackle deepfakes at the societal level, legislation, policy, action, and education are required (Shahzad et al., 2022). However, we contend that the issues surrounding deepfakes and the unique contexts of higher education institutions mean that stakeholders must begin making plans to tackle deepfake incidents as they become increasingly likely because of the rapid pace of development of the underlying technology.

Deepfakes are likely to become a dual-use technology (Gamage et al., 2022) with both positive and negative implications. At present, the major risks associated with deepfakes from an HEI perspective are their possible use for cyberbullying, their use to engage in academic dishonesty, their potential to spread disinformation, and sow social division. If these concerns can be proactively addressed, deepfake technology may present exciting opportunities for universities to create more engaging, inclusive, and accessible educational content and experiences. However, realising these benefits will require robust interdisciplinary research to better understand the pedagogical impacts, limitations, and ethical challenges.

To address these challenges, strategies should include augmenting cyberbullying protocols, adopting educational interventions, developing crisis management plans, and engaging in academic research to deepen our understanding of the impact of deepfakes in higher education and society at large. Research areas which we believe would support in a greater understanding of the risks, benefits, and impacts of deepfake technology within the HE landscape include

1. Examining faculty and student perceptions and vulnerabilities to deepfakes across various cultural contexts, to establish baselines of knowledge and identifying unique risks faced by marginalised groups.
2. Evaluating the effectiveness of different pedagogical strategies for building deepfake literacy and resilience among university students and staff and establishing best practices for classroom learning and campus awareness campaigns.
3. Conducting case studies of institutions that have dealt with deepfake-related incidents to inform the development of proactive and responsive protocols, policy approaches, and governance frameworks.
4. Exploring the ethical dimensions and potential beneficial applications of synthetic media production in teaching and learning, such as immersive simulations or low-cost video production, while developing frameworks to ensure responsible implementation.
5. Longitudinal and interdisciplinary research is needed to understand the long-range effects of deepfakes on core aspects of higher education, such as learning outcomes, academic integrity, and campus discourse, and to examine how deepfakes intersect with other key issues shaping the future of the sector.
6. Investigating the specific impacts on individuals who are victimised by malicious deepfakes and developing evidence-based resources and procedures to support those affected and prevent further harm.

By pursuing this research agenda, the academic community can play a key role in shaping the understanding of, and responses to the challenges and opportunities presented by deepfakes to both the HE sector and society. Through interdisciplinary scholarship, HEIs can develop evidence-based policies, practices, and pedagogies needed to mitigate the risks of synthetic media while harnessing any potential benefits for teaching, learning, and operations. As the pace of technological change in this area continues to accelerate, it is crucial that universities remain at the forefront of research and innovation to ensure that the power of deepfakes is wielded responsibly in the service of education and the public good.

**Acknowledgements**

Thank you to Leon Furze for providing additional insight into the topic and for the identification of additional literature sources.

**AI Usage Disclaimer**

This study used Generative AI tools (Claude 3 Opus) for draft text creation, revision, and editorial purposes throughout the production of the manuscript. The authors reviewed, edited, and take responsibility for all outputs of the tools used in this study.